\documentclass[aps,prl,reprint,superscriptaddress,amsmath,amssymb]{revtex4-2}

\usepackage{dcolumn}
\usepackage{bm}
\usepackage{graphicx}
\usepackage{amssymb}
\usepackage{amsmath, amscd}  
\newcommand{\R}{\mathbb{R}}

\newenvironment{itquote}
  {\begin{quote}\itshape}
  {\end{quote}\ignorespacesafterend}

\begin{document}

\title{Emergence of Classical Dynamics from a Random Matrix Schr{\"o}dinger Model}

\author{Alexey A. Kryukov}
\affiliation{Department of Mathematics \& Natural Sciences, University of Wisconsin-Milwaukee, USA}
\email{kryukov@uwm.edu}

\begin{abstract}
The Newtonian motion of a macroscopic particle is derived from the linear Schr{\"o}dinger equation with a Hamiltonian consisting of the free-particle term and a random Hamiltonian drawn from the Gaussian Unitary Ensemble. The random term models interaction with the environment. We show that the parameters governing the resulting state-space random walk, together with the treatment of experimentally indistinguishable states as equivalence classes, explain the contrasting behavior of microscopic and macroscopic systems. The analysis extends previous work deriving the Born rule for microscopic particles when the free-particle term is negligible.
\end{abstract}

\maketitle

\section{Motivation and Overview}

 It is widely asserted that linear dynamics cannot produce state reduction in quantum theory, which motivates the introduction of nonlinear collapse models. Such models, however, must reconcile linear unitary evolution with nonlinear state reduction while remaining consistent with stringent laboratory and cosmological bounds on collapse-induced noise \cite{Donadi2021,Piscicchia2024,Bilardello2016,Vinante2015,Carlesso2016,Altamura2025,Adler2007,AdlerBassi2009,IGMHeating2017}. These constraints exclude most parameter ranges, leaving only a narrow window of admissible values.

Without contradicting the standard proofs that rule out linear dynamics under certain assumptions, it was shown in \cite{KryukovPhysicsA} that these assumptions need not apply in a geometry based on equivalence classes of states. In that work, the equivalence classes represent states that are indistinguishable by a detector due to the finite resolution of measuring devices. Since unitary operators can ``squeeze" states into a given equivalence class, this framework naturally leads to a linear stochastic model of state reduction applicable to both classical and quantum measurements. In the classical regime, the model predicts a normal distribution for the position of a measured particle, while in the quantum regime it reproduces the Born rule.

Here, we extend these results by deriving Newtonian dynamics for macroscopic bodies from free Schr\"odinger evolution combined with a linear stochastic process on state space. We begin with a review of the previously developed formalism required for this derivation. In particular, the geometry of the embedding of Euclidean space and classical phase space into quantum state space, which connects Schr\"odinger and Newtonian dynamics, was developed in \cite{KryukovPhysicsA,Kryukov2020,KryukovNew}. The relationship between Poisson brackets and commutators of canonical observables that follows from this geometry was established in \cite{KryukovNew}. The embedding also yields a relation between commutators and the curvature of state space, discussed in \cite{Kryukov2020, KryukovDICE24} and, in a more general mathematical context, in \cite{KryukovMath}. The connection between the normal probability distribution and Born-rule statistics arising from this embedding was developed in \cite{KryukovPhysicsA} and \cite{Kryukov2020}.

The collapse model based on random Hamiltonians drawn independently in time from the Gaussian Unitary Ensemble, formulated as conjecture {\bf (RM)}, was introduced in \cite{KryukovPhysicsA,Kryukov2020,KryukovNew}. Equivalence classes of detector-indistinguishable states, together with numerical simulations confirming the emergence of Born-rule statistics under the {\bf (RM)} dynamics, were presented in \cite{KryukovPhysicsA}.

After briefly reviewing these results, we discuss the physical motivation for conjecture {\bf (RM)} through its analogy with Brownian motion and its connection with random-matrix ideas in quantum theory, and provide supporting numerical estimates. We then identify the conditions under which the free Schr\"odinger evolution and the evolution generated by the {\bf (RM)} Hamiltonian can be treated separately, confirming this numerically under physically reasonable assumptions. Finally, we show how this separation, combined with a physically justified choice of random-walk parameters in {\bf (RM)}, yields Newtonian motion for macroscopic objects and clarifies the distinction between microscopic and macroscopic regimes within a single framework. We also compare the present framework with existing approaches. The Supplement provides details on equivalence classes of states, the induced geometry of the corresponding manifolds, and the foliation structure used in the derivation of the Born rule.

We emphasize that, despite the physical motivation noted above, conjecture {\bf (RM)} remains exactly what it is claimed to be: a conjecture. Still, its consequences are mathematically precise and physically nontrivial. To our knowledge, no previous collapse-type or measurement model has shown simultaneously, within a single framework with physically justified parameters, (i) strict unitarity at the fundamental level, (ii) single outcomes distributed according to the Born rule, and (iii) recovery of classical motion. Moreover, the conjecture isolates a specific structural property, namely isotropic random-matrix effective dynamics, whose validity can in principle be investigated experimentally.

\section{Dynamics on state space}

Let \(M^{\sigma}_{3,3} \subset \mathbb{CP}^{L_2}\) denote the set of states represented by functions in \(L_2(\mathbb R^3)\) of the form
\begin{equation}
\label{phii}
\varphi({\bf x}) = r_{{\bf a}, \sigma}({\bf x}) \, e^{i{\bf p} \cdot {\bf x} / \hbar}.
\end{equation}
Here
\begin{equation}
\label{sigmao}
r_{{\bf a}, \sigma}({\bf x}) = \sigma^{-\frac{3}{2}} \, r\!\left( \frac{{\bf x} - {\bf a}}{\sigma} \right),
\end{equation}
where \(\sigma\) is a small positive parameter and \(r \in L_2(\mathbb R^3)\) is a fixed, normalized, real-valued, \(C^2\)-function with variance equal to \(1\). For \(\sigma \to 0\), one has \(r_{{\bf a}, \sigma}^2({\bf x}) \to \delta^3({\bf x}-{\bf a})\) \cite{delta}. Gaussian functions
\begin{equation}
\label{g}
g_{{\bf a}, \sigma}({\bf x}) = \left( \frac{1}{2\pi\sigma^{2}} \right)^{3/4} \exp\!\left[ -\frac{({\bf x} - {\bf a})^{2}}{4\sigma^{2}} \right]
\end{equation}
provide a convenient example, although the construction does not depend on the particular functional form of the representatives. 

So far, \(\sigma\) has been introduced as a free mathematical parameter controlling localization. Physically, however, it is natural to relate \(\sigma\) to the resolution of position-measuring devices. The resolution of a measuring device determines which states can be experimentally distinguished and thereby leads to equivalence classes of states that are indistinguishable to that device. In this interpretation, \(M^{\sigma}_{3,3}\) may be defined as the set of equivalence classes formed from states whose position standard deviation does not exceed \(\sigma\). This more general formulation will be introduced below (see also \cite{KryukovPhysicsA}).

The inclusion of  $M^{\sigma}_{3,3}$ into  $\mathbb{CP}^{L_2}$, furnished with the Fubini-Study metric, induces
the differentiable structure and the Riemannian metric on  $M^{\sigma}_{3,3}$.
With an appropriate choice of units, the map
\begin{equation}
\label{Omega}
\Omega: ({\bf a}, {\bf p}) \longmapsto r_{{\bf a}, \sigma} \, e^{i{\bf p} \cdot {\bf x} / \hbar}
\end{equation}
is an isometry between the Euclidean space $\mathbb{R}^3 \times \mathbb{R}^3$ and the Riemannian manifold $M^{\sigma}_{3,3}$.  
Moreover, a linear structure on $M^{\sigma}_{3,3}$ can be induced via $\Omega$ from the linear structure on $\mathbb{R}^3 \times \mathbb{R}^3$.  

Variation of the action functional 
\begin{equation}
\label{SS}
S[\varphi] = \int \overline{\varphi}({\bf x}, t) \left[i\hbar \frac{\partial}{\partial t} - {\widehat h} \right] \varphi({\bf x}, t) \, d^3{\bf x} \, dt,
\end{equation}
where the Hamiltonian ${\widehat h} $ is given by
\begin{equation}
\label{h}
{\widehat h} = -\frac{\hbar^{2}}{2m} \Delta + {\widehat V}({\bf x}, t) 
\end{equation}
with respect to $\varphi$, yields the Schr{\"o}dinger equation for a particle. 
By constraining \(\varphi\) in (\ref{SS}) to lie on the manifold \(M^{\sigma}_{3,3}\) and assuming continuity of the potential, the action functional takes the classical form
\begin{equation}
\label{SSS}
S = \int \left[ {\bf p} \cdot \frac{d{\bf a}}{dt} - h({\bf p}, {\bf a}, t) \right] dt,
\end{equation}
where
\begin{equation}
\label{h1}
h({\bf p}, {\bf a}, t) = \frac{{\bf p}^2}{2m} + V({\bf a}, t)
\end{equation}
is the Hamiltonian function of the system.  
Constrained variation yields the Newton equations. 

To see why this restriction gives the classical action, note that for the restricted ansatz
\(\varphi_{{\bf a},{\bf p}}({\bf x})=r_{{\bf a},\sigma}({\bf x})e^{i{\bf p}\cdot{\bf x}/\hbar}\), the non-potential terms in the Schr\"odinger action already reduce to the classical form: the term \(i\hbar\langle \varphi,\partial\varphi/\partial t\rangle\) equals \({\bf p}\cdot d{\bf a}/dt\) up to a total derivative, and the kinetic term equals \({\bf p}^2/2m\) up to an additive \(\sigma\)-dependent constant independent of \({\bf a},{\bf p}\). Thus the only genuine approximation in passing from the restricted Schr\"odinger action to the classical action comes from the potential term
\begin{equation}
\int_{\mathbb R^3} V({\bf x},t)\,r^2_{{\bf a},\sigma}({\bf x})\,d^3{\bf x},
\end{equation}
which converges to \(V({\bf a},t)\) as \(\sigma\to 0\) under mere continuity of \(V\) at \({\bf a}\).

For a general state of the form
\begin{equation}
\varphi({\bf x},t)=r_{{\bf a},\sigma}({\bf x})\,e^{i\Theta({\bf x},t)},
\end{equation}
assuming differentiability of the phase \(\Theta\) at \({\bf a}\), one may write
\begin{equation}
\Theta({\bf x},t)
=
\Theta_0(t)
+
\frac{1}{\hbar}{\bf p}(t)\cdot({\bf x}-{\bf a}(t))
+
R({\bf x},t),
\end{equation}
where \(R({\bf x},t)=o(|{\bf x}-{\bf a}|)\).
Since the state is localized on the scale \(\sigma\), one has \(R=o(\sigma)\) on the localization region. It follows that, for sufficiently small \(\sigma >0\), the contribution of the nonlinear part of the phase to the restricted Schr\"odinger action tends to zero with \(\sigma\), so that the action differs from the one obtained for a linear phase by a term vanishing as \(\sigma \to 0\). This observation will be important below when \(M_{3,3}^\sigma\) is reformulated in terms of equivalence classes of localized states.

From the obtained relation between the Schr\"odinger and classical actions, it follows that the parameters \(({\bf a},{\bf p})\) play the role of classical position and momentum on \(M_{3,3}^\sigma\). In particular, the parameter \({\bf p}\) is associated with the linear phase factor \(e^{i{\bf p}\cdot {\bf x}/\hbar}\) and determines the group velocity of the packet, which is the physically relevant quantity for the transition to classical behavior. Thus the construction does not require sharply defined momentum eigenstates. Rather, it singles out a phase-space manifold of localized states whose induced Fubini--Study geometry is Euclidean and whose tangent Schr\"odinger evolution reproduces Newtonian dynamics in the variables \(({\bf a},{\bf p})\).

The construction extends to many-body systems. In particular, a two-particle system with Hamiltonian 
\begin{equation}
\label{twoh}
{\widehat h} = -\frac{\hbar^{2}}{2m_1} \Delta_1 -\frac{\hbar^{2}}{2m_2} \Delta_2 +  {\widehat V}({\bf x}_1, {\bf x}_2, t)
\end{equation} 
whose state is constrained to the manifold $M^{\sigma}_{3,3} \otimes M^{\sigma}_{3,3}$ evolves according to Newtonian dynamics.

Using the Gaussian representative \(r_{{\bf a},\sigma}=g_{{\bf a},\sigma}\) until equivalence classes of states are introduced, one can demonstrate directly from the Schr\"odinger equation the role played by the constraint to $M^{\sigma}_{3,3}$.
As shown in \cite{KryukovNew}, the Schr\"odinger velocity of a state $\varphi \in M^{\sigma}_{3,3}$ decomposes into three orthogonal parts: two tangent to $M^{\sigma}_{3,3}$, corresponding to the classical velocity and acceleration, and one normal to the manifold that describes wave-packet spreading. The squared norms of these components add up to
\begin{equation}
\label{decomposition}
\left\| \frac{d\varphi}{dt} \right\|^{2}_{FS}
= \frac{{\bf v}^{2}}{4\sigma^{2}}
+ \frac{m^{2}{\bf w}^{2}\sigma^{2}}{\hbar^{2}}
+ \frac{\hbar^{2}}{32\sigma^{4}m^{2}},
\end{equation}
where ${\bf v}$ is the classical velocity and 
${\bf w}=-\nabla V/m$ the classical acceleration.
Constraining the motion to $M^{\sigma}_{3,3}$ suppresses the normal (spreading) component of
\begin{equation}
\frac{d\varphi}{dt}=-\frac{i}{\hbar}\widehat h\varphi,
\end{equation}
reducing commutators to Poisson brackets and yielding Newtonian dynamics
\cite{KryukovNew}.

The relation between the action functionals (\ref{SS}) and (\ref{SSS}) identifies Newtonian particles with quantum systems whose states are constrained to the manifold $M^{\sigma}_{3,3}$ for suitable $\sigma$. The isometry $\Omega$ then provides a direct identification between the Euclidean phase space $\mathbb{R}^3 \times \mathbb{R}^3$ of a Newtonian particle and the manifold of states $M^{\sigma}_{3,3}$. Its restriction
\begin{equation}
\omega:\,{\bf a}\mapsto g_{{\bf a},\sigma}
\end{equation}
is an isometry between the Euclidean position space $\mathbb{R}^3$ and the submanifold $M^{\sigma}_3\subset\mathbb{CP}^{L_2}$ of states $g_{{\bf a},\sigma}$  \cite{Kryukov2020}. 
This isometry is captured by the relation
\begin{equation}
\label{mainO}
e^{-\frac{({\bf a}-{\bf b})^{2}}{4\sigma^{2}}}
= \cos^{2} \rho( g_{{\bf a},\sigma}, g_{{\bf b},\sigma}),
\end{equation}
which relates the Euclidean distance $||{\bf a}-{\bf b}||$ with the Fubini-Study distance $\rho( g_{{\bf a},\sigma}, g_{{\bf b},\sigma})$ between the corresponding Gaussian states. For states including momenta, $\varphi=g_{{\bf a},\sigma}e^{i{\bf p}\cdot{\bf x}/\hbar}$ and $\psi=g_{{\bf b},\sigma}e^{i{\bf q}\cdot{\bf x}/\hbar}$, the analogous relation is
\begin{equation}
\label{mainOO}
e^{-\frac{({\bf a}-{\bf b})^{2}}{4\sigma^{2}}
-\frac{({\bf p}-{\bf q})^{2}}{\hbar^{2}/\sigma^{2}}}
= \cos^{2}\rho(\varphi,\psi).
\end{equation}

The special role of the submanifolds $M^{\sigma}_{3,3}$ and $M^{\sigma}_{3}$ is evident not only in establishing the connection between Schr\"odinger and Newtonian dynamics, but also in relating the canonical commutation relations to the curvature of the state space.
As a first step, the embedding of the Euclidean classical phase space into the curved state space $\mathbb{CP}^{L_2}$ naturally gives rise to nontrivial position-momentum commutators. For states $\varphi\in M^{\sigma}_{3,3}$, the position operator acts as the generator of momentum displacements,
\begin{equation}
\label{vector}
{\widehat {\bf x}}\varphi=-i\hbar \nabla_{\bf p}\varphi.
\end{equation}
The vector fields $-\hbar \nabla_{\bf p}\varphi$ and $-\hbar \nabla_{\bf x}\varphi$ commute, and their integral curves furnish orthogonal coordinates on $M^{\sigma}_{3,3}$, confirming its identification with the Euclidean phase space $\R^3 \times \R^3$. 

Because states in \(M^{\sigma}_{3,3}\) form a complete set in the Hilbert space, the vector field \(-\hbar \nabla_{\bf p}\varphi\) extends uniquely to a linear vector field on the full space, which by (\ref{vector}) is precisely \(-i\widehat{\bf x}\varphi\) \cite{KryukovPhysLett}. Although the Lie bracket of \(-\hbar \nabla_{\bf p}\varphi\) and \(-\hbar \nabla_{\bf x}\varphi\) vanishes on \(M^{\sigma}_{3,3}\), the bracket of their extensions, \(-i\widehat{\bf p}\varphi\) and \(-i\widehat{\bf x}\varphi\), tangent to the unit sphere \(\mathbb{S}^{L_2}\) or pushed down to \(\mathbb{CP}^{L_2}\), yields
\begin{equation}
\label{CCR}
[\widehat{\bf x},\widehat{\bf p}]\,\varphi=i\hbar\,\varphi.
\end{equation}
Since \(\varphi\) is now an arbitrary normalized state, equation (\ref{CCR}) is simply the operator equality. The derivation therefore provides, within the proposed framework, the classical geometric origin of the operators satisfying the canonical commutation relation, as unique extensions of the corresponding classical quantities.

Furthermore, unlike the submanifold \(M^{\sigma}_{3,3}\), the full state space \(\mathbb{CP}^{L_2}\) has nontrivial curvature. The sectional curvature in the plane spanned by the tangent vectors \(-i\widehat{\mathbf{p}}\varphi\) and \(-i\widehat{\mathbf{x}}\varphi\) can be expressed through their Lie bracket, that is, through the commutator \([\widehat{\mathbf{x}},\widehat{\mathbf{p}}]\). The curvature is independent of \(\varphi\) and corresponds to the sphere \(\mathbb{S}^2=\mathbb{CP}^1\) of radius \(\hbar/2\) in that section \cite{Kryukov2020}. Thus the classical phase space appears as a flat submanifold embedded in a Planck-scale curved state space. The ``quantumness'' of the microworld arises from extending classicality from the manifolds \(M^{\sigma}_3=\mathbb{R}^3\) and \(M^{\sigma}_{3,3}=\mathbb{R}^3\times\mathbb{R}^3\), which ``wind'' through the infinite dimensions of \(\mathbb{CP}^{L_2}\), to the entire state space. The nontrivial canonical commutation relation between \(\widehat{\mathbf{x}}\) and \(\widehat{\mathbf{p}}\) thus reflects the nonvanishing sectional curvature of \(\mathbb{CP}^{L_2}\).

It turns out that the manifolds $M^{\sigma}_{3,3}$ and $M^{\sigma}_{3}$ are also essential for establishing a connection between measurement processes in classical and quantum systems. The familiar normal distribution of position outcomes for a classical particle arises when the measurement is modeled as a random walk of its position in $\mathbb{R}^3$ during the measurement interval, approximating Brownian motion. Such a model
is physically well motivated, since measurement errors result from the cumulative effect of many small fluctuations produced by interactions between the particle, the
measuring device, and the surrounding environment.

The following proposition defines an extension of this walk from $M^{\sigma}_3 = \mathbb{R}^3$ to the full state space $\mathbb{CP}^{L_2}$. We will show that the proposition parallels Einstein's assumptions for Brownian motion \cite{Ein}, to which the process reduces when constrained to \(M^{\sigma}_{3}\), and yields a model of position measurement for microscopic particles:

\begin{itquote}{\bf (RM)}
The dynamics of a particle's state under position measurement can be modeled as a random walk in the space of states. In the absence of drift, each step is generated by the Schr{\"o}dinger equation with a Hamiltonian drawn independently at each instant from the Gaussian Unitary Ensemble (GUE).
\end{itquote}
Here {\bf (RM)} denotes ``random matrices". Physically, such Hamiltonians may arise from the complicated and rapidly fluctuating interaction between the particle and its measuring apparatus or environment, reminiscent of Wigner's approach to complex spectra \cite{Wigner} and of the Bohigas-Giannoni-Schmit conjecture \cite{BGS}, here applied in a time-dependent setting. 

Although {\bf (RM)} is a conjecture, the forthcoming estimates provide support for it. Conjecture {\bf (RM)} yields precisely the features expected of a quantum measurement: the previously established uniqueness of the measurement outcome and validity of the Born rule, together with the transition to Newtonian dynamics and classical measurement demonstrated in this paper.

A small step in the state's random walk, driven by the Hamiltonian in {\bf (RM)}, is represented by a random vector in the tangent space to $\mathbb{CP}^{L_2}$. The distribution of such steps is normal, homogeneous, and isotropic \cite{KryukovNew}, implying that the transition probability between two states depends only on their Fubini-Study distance.
Indeed, since the GUE is invariant under unitary transformations and the action of the unitary group on $\mathbb{CP}^{L_2}$ is transitive, the transition probability is identical for any pair of states separated by the same Fubini-Study distance. Note that the action of the orthogonal group is not transitive on $\mathbb{CP}^{L_2}$. In particular, replacing the GUE with the GOE would not yield an analogous result.

When the steps are constrained to $M^{\sigma}_3$, the walk reduces to a walk with Gaussian steps on $\R^3$, which yields Brownian motion on $\mathbb{R}^3$, and the transition probability follows a normal distribution. 
 Without this constraint, the same isotropic process yields the Born rule \cite{KryukovPhysicsA,Kryukov2020,KryukovDICE24,KryukovNew}.
Under appropriate conditions, the connection between the normal probability distribution on the classical submanifold and the Born rule on state space can also be derived directly from (\ref{mainO}). This establishes a dynamical link between classical and quantum measurements, placing them on an equal footing.

Since Newtonian motion and the model of macroscopic position measurement are obtained by constraining the Schr\"odinger evolution and the random walk in {\bf (RM)} to \(M^{\sigma}_{3,3}\) and \(M^{\sigma}_3\), a dynamical explanation of this constraint is required. We show that an appropriate choice of the time-step and step-size parameters in {\bf (RM)} enforces this constraint and thereby distinguishes microscopic from macroscopic behavior. As noted above, the finite resolution of measuring devices implies that sufficiently localized states must be treated in terms of equivalence classes rather than individual states. We now introduce these equivalence classes.

Restricting for simplicity to one spatial dimension with state spaces $L_{2}(\mathbb{R})$ and $\mathbb{CP}^{L_{2}}$, the classical Euclidean
submanifold $M^{\sigma}_{1}=\mathbb{R}$ is represented, in particular, by the Gaussian states
\begin{equation}
\label{g1}
g_{a,\sigma}(z)
=\left(\frac{1}{2\pi\sigma^{2}}\right)^{1/4}
\exp\!\left[-\frac{(z-a)^{2}}{4\sigma^{2}}\right].
\end{equation}
Among all functions in $L_{2}(\mathbb{R})$ with finite position expectation $\mu_z$ and standard deviation $\delta_{z} \le \sigma$, the equivalence class $\{g_{c}\}$, interpreted as a physical eigenstate of position, consists of those with expectation value $\mu_{z}=c$.
For a measuring device whose position resolution is no finer than \(\sigma\), all states in a given equivalence class yield experimentally indistinguishable probabilities for detecting the particle within an interval of resolution size centered at \(\mu_z\).

The Fubini-Study distance between a state $\varphi \in L_2(\R)$ and an equivalence class $\{g_c\}$ is defined by
\begin{equation}
\label{dist}
\rho(\varphi, \{g_c\})=\inf_{\psi \in \{g_c\}} \rho(\varphi, \psi),
\end{equation}
where $\rho(\varphi,\psi)$ denotes the Fubini-Study distance between states. 
A state $\varphi$ reaches the physical eigenstate $\{g_c\}$ precisely when $\rho(\varphi,\{g_c\})=0$. 
The distance between nearby equivalence classes $\{g_c\}$ and $\{g_d\}$ is defined similarly:
\begin{equation}
\label{dist1}
\rho(\{g_c\}, \{g_d\})=\inf_{\varphi \in \{g_c\}} \rho(\varphi, \{g_d\}).
\end{equation}
For $d$ sufficiently close to $c$, this distance agrees with the distance between the Gaussian representatives $g_{c,\sigma}$ and $g_{d,\sigma}$ in $M^{\sigma}_1$, up to
$o(|c-d|/\sigma)$ corrections. In particular, these representatives determine the induced line element on $\widetilde M^{\sigma}_1$. Hence the manifold $\widetilde M^{\sigma}_1$ of equivalence classes $\{g_c\}$, globally parametrized by $c \in {\mathbb{R}}$ and equipped with the metric defined by (\ref{dist1}), is isometric to $M^{\sigma}_1={\mathbb{R}}$.

The two-dimensional manifold $M^{\sigma}_{1,1}$ is defined analogously to $M^{\sigma}_{3,3}$. The manifold ${\widetilde M^{\sigma}_{1,1}}$ of equivalence classes is obtained by augmenting the equivalence classes of real-valued functions satisfying $\delta_z \le \sigma$ with a factor $e^{ipz/\hbar}$.
The distance between a state and an equivalence class in ${\widetilde M^{\sigma}_{1,1}}$ is defined as in (\ref{dist}). The metric on  ${\widetilde M^{\sigma}_{1,1}}$ is defined using the Gaussian representatives $g_{c,\sigma}e^{ipz/\hbar}$. The isometry between $M^{\sigma}_{1,1}$, ${\widetilde M^{\sigma}_{1,1}}$, and $\mathbb{R}^2$ is established using (\ref{mainOO}). See the Supplement for details.

As shown in \cite{KryukovPhysicsA}, in this framework the position of a classical particle is specified by an equivalence class of states labeled by the expectation value \(\mu_z\) and width \(\delta_z\). The sets of states satisfying \(\mu_z=\tau\) and \(\delta_z=\lambda\) are the level sets of the map
\begin{equation}
F(\varphi)=(\mu_z,\delta_z),
\end{equation}
and form the leaves of a codimension-two foliation of state space. In \cite{KryukovPhysicsA} and in the Supplement, we show that by translating and scaling any suitable initial state \(\varphi\), one obtains a two-dimensional submanifold \(M_\varphi\subset\mathbb{CP}^{L_2}\), equipped with the induced metric, on which \(\tau=\mu_z\) and \(s=\ln(\delta_z/\sigma)\) serve as orthogonal coordinates. Thus \(M_\varphi= \mathbb{R}^2\), and points in \(M_\varphi\) parameterize the leaves of the foliation. State reduction is then described by a stochastic process on \(\mathbb{R}^2\).

The random walk in {\bf (RM)} on \(\mathbb{CP}^{L_2}\) reduces to a Gaussian random walk on \(\mathbb{R}^2\). The component along \(\tau\) yields the normal distribution, which by isotropy extends uniquely to the Born rule on \(\mathbb{CP}^{L_2}\), while, by symmetry of the walk, the component along \(s\) gives probability approximately \(1/2\) that the state lies on \(\widetilde M^{\sigma}_{1}\), i.e., satisfies \(\delta_z \le \sigma\) \cite{KryukovPhysicsA}. This accounts for both the nonzero probability of reaching \(\widetilde M^{\sigma}_{1}\) and the emergence of the Born rule, and it plays a direct role in deriving the Newtonian behavior of macroscopic bodies from {\bf (RM)}.

\section{Main Result: Emergence of Newtonian Dynamics for Macroscopic Bodies}

Newtonian motion and the state of rest of macroscopic bodies presuppose experimentally accessible knowledge of their positions and velocities at any given moment. Such information arises not only from explicit observation, e.g., illumination and detection of scattered light, but also from continual interaction with the environment, which records the body's position and momentum through the scattering of ambient particles and radiation. For Newtonian motion to hold without explicit reference to the environment, these interactions must be weak yet not absent; otherwise neither position nor momentum would be well defined, and the notion of Newtonian motion would lose its operational meaning.

This effective ``continuous" measurement, together with the small diffusion coefficient of the associated Brownian motion, allows the body's Newtonian trajectory to remain well defined over long times without significant growth of positional uncertainty. Environmental interactions cause the position distribution to spread, but the acquisition of new positional information repeatedly contracts it. This alternating cycle of spreading and contraction stabilizes the observed Newtonian trajectory. Although the normal distribution permits rare large deviations from the Newtonian path, such events are extremely unlikely and lie within the tolerance of classical measurements.

Let us show that, with appropriate parameter choices, the walk in {\bf (RM)}, combined with free Schr\"odinger evolution, reproduces this behavior on the full state space, extending the classical dynamics defined on \(M^{\sigma}_3 = \mathbb{R}^3\) (or \(M^{\sigma}_1 = \mathbb{R}\)) and simultaneously yielding Newtonian motion for macroscopic bodies. 
Mathematically, the process we aim to derive is a random walk with intermittent conditioning. The {\bf (RM)} term produces isotropic random increments in projective state space, while the Schr\"odinger flow contributes Newtonian tangent drift whenever the state lies sufficiently close to \(\widetilde M_{1,1}^{\sigma}\). When the path reaches the \(\sigma\)-localized sector \(\widetilde M_{1}^{\sigma}\), the coordinate \(a\) is recorded and the subsequent walk is conditioned on the recorded value. The observed positions then form a sequence of conditional random variables centered on the Newtonian trajectory. The task is to show that physically reasonable choices of the time step and step variance allow the {\bf (RM)} and Schr\"odinger contributions to be separated and keep these conditional distributions narrow on the resolution scale \(\sigma\).

We first show that, during each short interaction window, the free evolution contributes only an observationally negligible tangent drift together with negligible orthogonal spreading, so that on the dynamically relevant sector its commutator with the {\bf (RM)} term is negligible.
We assume that the external potential \(V\) is differentiable in a neighborhood of \(a\). Then
$
V(x)=V(a)+\nabla V(a)\cdot(x-a)+o(|x-a|),
$
so on a localization region of size \(\sigma\) the potential is linear up to an \(o(\sigma)\) correction, which is sufficient for the tangent component of the free Schr\"odinger velocity on \(M_{1,1}^{\sigma}\) to reproduce the corresponding Newtonian drift step.

Let the manifold \(M_{1,1}^\sigma\) consist of states of a macroscopic point particle at position resolution \(\sigma\). The projected Schr\"odinger velocity of a state \(\psi=\psi_{a,p}^\sigma \in M_{1,1}^\sigma\) admits the orthogonal decomposition
\begin{equation}
\label{COMdecomposition}
-\frac{i}{\hbar}\widehat h\psi
=
\frac{da}{dt}\,\frac{\partial \psi}{\partial a}
+
\frac{dp}{dt}\,\frac{\partial \psi}{\partial p}
+
X_\perp,
\end{equation}
where the first two terms are tangent to \(M_{1,1}^\sigma\) and \(X_\perp\) is orthogonal to \(M_{1,1}^\sigma\). Here
\begin{equation}
\frac{da}{dt}=\frac{p}{M}, \qquad \frac{dp}{dt}=-\frac{dV}{da},
\end{equation}
and
\begin{equation}
\|X_\perp\| \sim \frac{\hbar}{M\sigma^2}.
\end{equation}
This is the one-dimensional specialization of the decomposition \eqref{decomposition} of the Schr\"odinger velocity. For states on \(M_{1,1}^\sigma\), the free-evolution velocity thus decomposes into a tangent component generating classical drift and an orthogonal component responsible for spreading.

 A macroscopic body is continuously monitored by its environment through scattering of air molecules and ambient radiation. Each such scattering event is localized in time and may be treated as a short interaction episode (``kick") during which position information is transferred to environmental degrees of freedom.
Let us estimate the size of these contributions during a single environmental collision.

\paragraph{Tangent displacement during one collision.}

For air at room temperature, the typical thermal velocity of molecules is
\begin{equation}
v_{\mathrm{th}} \sim 5\times10^2\,\mathrm{m/s}.
\end{equation}
Taking a conservative interaction range of molecular scale
\begin{equation}
\ell \sim 1\,\mathrm{nm} = 10^{-9}\,\mathrm{m},
\end{equation}
the duration of a collision is
\begin{equation}
\tau_{\mathrm{air}} \sim \frac{\ell}{v_{\mathrm{th}}}
\sim
2\times10^{-12}\,\mathrm{s}.
\end{equation}
For photons, $v_{\mathrm{rel}} = c$. Even taking a conservative scale
\begin{equation}
\ell \sim 1\,\mu\mathrm{m} = 10^{-6}\,\mathrm{m},
\end{equation}
one obtains
\begin{equation}
\tau_{\gamma} \sim \frac{\ell}{c}
\sim
3\times10^{-15}\,\mathrm{s}.
\end{equation}
Thus realistic environmental interaction windows satisfy
\begin{equation}
\tau \lesssim 10^{-12}\,\mathrm{s}
\quad
(\text{air}),
\qquad
\tau \lesssim 10^{-15}\,\mathrm{s}
\quad
(\text{radiation}).
\end{equation}
Take
\begin{equation}
\sigma = 1\,\mu\mathrm{m} = 10^{-6}\,\mathrm{m}.
\end{equation}
For a macroscopic body moving with $v \lesssim 1\,\mathrm{m/s}$,
\begin{equation}
\Delta a \sim v\tau.
\end{equation}
For air:
\begin{equation}
\Delta a_{\mathrm{air}}
\sim
(1)(2\times10^{-12})
=
2\times10^{-12}\,\mathrm{m}.
\end{equation}
For photons:
\begin{equation}
\Delta a_{\gamma}
\sim
3\times10^{-15}\,\mathrm{m}.
\end{equation}
Relative to $\sigma$,
\begin{equation}
\frac{\Delta a}{\sigma}
\sim
\begin{cases}
2\times10^{-6} & \text{(air)},\\[4pt]
3\times10^{-9} & \text{(radiation)}.
\end{cases}
\end{equation}
Hence
\begin{equation}
\frac{v\tau}{\sigma} \ll 1
\end{equation}
by six to nine orders of magnitude. The classical (tangent) drift generated by ${\widehat h}$ is therefore dynamically invisible at $\sigma$-resolution during each collision.

\paragraph{Orthogonal spreading during one collision.}

The spreading time scale is
\begin{equation}
T_{\mathrm{spr}} = \frac{M\sigma^2}{\hbar}.
\end{equation}
For example, even for a small macroscopic mass
\begin{equation}
M = 10^{-6}\,\mathrm{kg},
\qquad
\sigma = 10^{-6}\,\mathrm{m},
\end{equation}
\begin{equation}
T_{\mathrm{spr}}
=
\frac{10^{-6}\cdot10^{-12}}{10^{-34}}
\sim
10^{16}\,\mathrm{s}.
\end{equation}
Thus
\begin{equation}
\frac{\tau}{T_{\mathrm{spr}}}
\lesssim
\begin{cases}
10^{-28} & \text{(air)},\\[4pt]
10^{-31} & \text{(radiation)}.
\end{cases}
\end{equation}
Therefore
\begin{equation}
\|X_\perp\|\tau \ll 1
\end{equation}
is exceptionally small and the orthogonal component of the free Schr\"odinger velocity cannot accumulate during a collision window.

\paragraph*{Effective commutation on the $\sigma$-resolved sector.}

Let \(P_\sigma\) denote the projection onto the \(\sigma\)-localized tube around the manifold \(M_{1,1}^\sigma\) of states in the Hilbert space. Thus \(\psi\in\operatorname{Ran}(P_\sigma)\), where \(\operatorname{Ran}(P_\sigma)\) denotes the range of the operator \(P_\sigma\), means that \(\psi\) lies in a \(\sigma\)-neighborhood of \(M_{1,1}^\sigma\) in the Hilbert-space norm. Since we are working here with normalized representatives in Hilbert space rather than directly in projective space, the free evolution may include an overall phase factor \(e^{i\theta}\), where \(\theta\) is a real phase depending on the state and the short time interval \(\tau\). The above bounds imply that for \(\psi \in \operatorname{Ran}(P_\sigma)\),
\begin{equation}
e^{-\frac{i}{\hbar}\widehat h\tau}\psi
=
e^{i\theta}\psi
+
O\!\left(\frac{v\tau}{\sigma}\right)
+
O\!\left(\frac{\tau}{T_{\mathrm{spr}}}\right).
\end{equation}
Hence, on the \(\sigma\)-resolved sector,
\begin{equation}
\widehat h\psi
=
E_\psi \psi
+
O(\varepsilon),
\qquad
\varepsilon
\sim
\frac{v\tau}{\sigma}
+
\frac{\tau}{T_{\mathrm{spr}}},
\end{equation}
where \(E_\psi\) is the scalar by which \(\widehat h\) acts on \(\psi\) to leading order. For the numerical values above,
\begin{equation}
\varepsilon \lesssim 10^{-6}
\qquad
(\text{tangent and orthogonal contributions}).
\end{equation}
Let \(\widehat h_{\mathrm{RM}}\) denote the Hamiltonian in {\bf (RM)}. Then, for \(\psi \in \operatorname{Ran}(P_\sigma)\),
\begin{equation}
[\widehat h,\widehat h_{\mathrm{RM}}]\psi
=
E_\psi \widehat h_{\mathrm{RM}}\psi
-
\widehat h_{\mathrm{RM}}(E_\psi\psi)
+
O(\varepsilon)
=
O(\varepsilon),
\end{equation}
and consequently
\begin{equation}
P_\sigma[\widehat h,\widehat h_{\mathrm{RM}}]P_\sigma
=
O(\varepsilon),
\end{equation}
with \(\varepsilon \ll 1\) for realistic environmental parameters.


So, for macroscopic bodies at resolution \(\sigma=1\,\mu\mathrm{m}\), the following holds: (i) environmental interactions occur in extremely short windows (\(\tau\lesssim10^{-12}\,\mathrm{s}\)); (ii) during each such window, free evolution produces a \(\sigma\)-invisible tangent shift and negligible orthogonal spreading; (iii) consequently, on the dynamically relevant \(\sigma\)-localized sector, the action of the free Hamiltonian differs from a scalar action only by a very small error; (iv) therefore, the commutator \([\widehat h,\widehat h_{\mathrm{RM}}]\) vanishes to high accuracy on the states of interest; (v) the total evolution can thus be described as alternating free segments generating Newtonian tangent flow on \(\widetilde M_{1,1}^\sigma\) and short environmental ``kicks''.

Our next goal is to estimate the time step $dt$ and  Gaussian step variance $(dz)^2$ parameters of the random walk in {\bf (RM)} for macroscopic body.
The environmental interaction described above induces extremely frequent
scattering events. It is therefore natural to describe the dynamics
in a coarse-grained stochastic form, where each short time interval
$ dt$ contains many microscopic collisions.

\paragraph{Effective time step.}

For air at room temperature,
\begin{equation}
v_{\mathrm{th}} \sim 5\times10^2\,\mathrm{m/s},
\qquad
\ell \sim 1\,\mathrm{nm},
\end{equation}
so that the collision duration is
\begin{equation}
\tau_{\mathrm{air}}
\sim
\frac{\ell}{v_{\mathrm{th}}}
\sim
2\times10^{-12}\,\mathrm{s}.
\end{equation}
A natural coarse-grained time step is therefore
\begin{equation}
dt \sim \tau_{\mathrm{air}} \sim 10^{-12}\,\mathrm{s}.
\end{equation}
The molecular flux at standard conditions is
\begin{equation}
\Phi \sim \frac{n v_{\mathrm{th}}}{4}
\sim 3\times10^{27}\,\mathrm{m^{-2}s^{-1}},
\end{equation}
and for a body of radius $R$ with cross-sectional area
$A \sim \pi R^2$ the collision rate is
\begin{equation}
\Gamma_{\mathrm{air}} \sim \Phi A.
\end{equation}
For $R = 1\,\mathrm{mm}$,
\begin{equation}
\Gamma_{\mathrm{air}} \sim 10^{22}\,\mathrm{s^{-1}}.
\end{equation}
Hence during one coarse-grained interval $dt \sim 10^{-12}\,\mathrm{s}$,
the number of independent microscopic collisions is
\begin{equation}
N = \Gamma_{\mathrm{air}} dt \sim 10^{10} \gg 1.
\end{equation}
This is precisely the regime in which the net environmental effect
is well approximated by a diffusion process.

\paragraph{Step variance and diffusion coefficient.}

Let $\Delta p_k$ denote the momentum transfer from the $k$-th collision.
Each transfer has characteristic magnitude
\begin{equation}
|\Delta p_k| \sim p_{\mathrm{air}}
\sim m_{\mathrm{air}} v_{\mathrm{th}}
\sim 2.5\times10^{-23}\,\mathrm{kg\,m/s}.
\end{equation}
Over $N$ independent collisions within $dt$,
the total momentum increment satisfies
\begin{equation}
\mathbb{E}[\Delta p] = 0,
\qquad
\mathrm{Var}(\Delta p)
\sim
N p_{\mathrm{air}}^2.
\end{equation}
Thus
\begin{equation}
\mathrm{Var}(\Delta p)
\sim
\Gamma_{\mathrm{air}} dt \, p_{\mathrm{air}}^2.
\end{equation}
The associated momentum diffusion coefficient is
\begin{equation}
D_p \sim \Gamma_{\mathrm{air}} p_{\mathrm{air}}^2.
\end{equation}
The induced position diffusion coefficient is
\begin{equation}
D_a \sim \frac{D_p}{M^2}.
\end{equation}

\paragraph{Magnitude of stochastic corrections.}

For a macroscopic body of mass
\begin{equation}
M = 10^{-6}\,\mathrm{kg},
\end{equation}
the classical momentum at $v \sim 1\,\mathrm{m/s}$ is
\begin{equation}
p_{\mathrm{macro}} \sim 10^{-6}\,\mathrm{kg\,m/s}.
\end{equation}
Using $N \sim 10^{10}$,
\begin{equation}
\Delta p_{\mathrm{RM}}
\sim
p_{\mathrm{air}} \sqrt{N}
\sim
(2.5\times10^{-23})(10^5)
\sim
2.5\times10^{-18}\,\mathrm{kg\,m/s}.
\end{equation}
Hence
\begin{equation}
\frac{\Delta p_{\mathrm{RM}}}{p_{\mathrm{macro}}}
\sim 10^{-12}.
\end{equation}
Thus even though the environmental kicks are extremely frequent,
their cumulative effect on the macroscopic momentum over $dt$
is negligible compared to the deterministic classical value.

\paragraph{Stroboscopic Newtonian motion.}

On the manifold \(M_{1,1}^\sigma\),
the deterministic tangent flow generated by the Hamiltonian satisfies
\begin{equation}
\frac{da}{dt} = \frac{p}{M},
\qquad
\frac{dp}{dt} = -\frac{dV}{da}.
\end{equation}
The environmental {\bf (RM)}-based contribution produces stochastic increments
\begin{equation}
\Delta a_{\mathrm{RM}} \sim \xi_a \sqrt{dt},
\qquad
\Delta p_{\mathrm{RM}} \sim \xi_p \sqrt{dt},
\end{equation}
where \(\xi_a\) and \(\xi_p\) are random variables with zero mean and variances determined by \(D_a\) and \(D_p\), respectively.
Because
\begin{equation}
\mathbb{E}[\Delta p_{\mathrm{RM}}]=0,
\qquad
\mathbb{E}[\Delta a_{\mathrm{RM}}]=0,
\end{equation}
the mean motion satisfies the classical equations of motion,
while fluctuations accumulate only diffusively.

The orthogonal deviation from the classical manifold resulting from the random walk
may be described in terms of the variable
$s=\ln(\delta_z/\sigma)$, where the standard deviation $\delta_z$ measures the transverse distance from the state to $M_{1,1}^\sigma$.
Under the {\bf (RM)} dynamics with negligible drift in \(s\), the discrete-time process \(s_n\), evaluated at kick times \(t_n=n\,dt\), where \(n=0,1,2,\dots\), is a symmetric random walk. By the Sparre Andersen theorem, the probability that the process has not returned to \(s\le 0\) after \(n\) steps satisfies
\begin{equation}
\mathbb{P}(\tau>n)
=
\frac{\binom{2n}{n}}{2^{2n}}
\sim \frac{1}{\sqrt{\pi n}}.
\end{equation}
Hence, with probability $0.999968$ (four standard deviation contribution),
the state returns to the tubular neighborhood
$\delta_z\le \sigma$
within approximately
\begin{equation}
n \approx \frac{1}{\pi(1-0.999968)^2}
\approx 3.1\times 10^{8}
\quad \text{steps}.
\end{equation}
For the physically motivated coarse-grained step
$dt\sim 10^{-12}\,\mathrm s$,
this corresponds to a time
\begin{equation}
T_{0.999968} \approx 3\times 10^{-4}\,\mathrm s.
\end{equation}

Thus, with overwhelming probability,
the state under the transverse diffusion returns to $M^{\sigma}_{1,1}$
on a sub-millisecond time scale.
As shown earlier in the paper, when the state lies on $M^{\sigma}_{1,1}$ (or $\widetilde M^{\sigma}_{1,1}$),
its Schr\"odinger evolution coincides with Newtonian motion
on the classical manifold.
Environmental scattering then encodes the particle's position into the surroundings in an effectively classical manner.
This transfer of information reinitializes the coarse-grained diffusion.

Moreover, the short return time provides an upper bound on the classical displacement accumulated during a renewal cycle. The stochastic contribution to the classical position satisfies
\begin{equation}
\Delta a_{\mathrm{RM}} \sim \sqrt{D_a\,T},
\qquad
D_a = \frac{D_p}{M^2},
\end{equation}
where $D_p$ arises from environmental scattering. For representative macroscopic parameters (e.g. air at room temperature and \ $M\sim10^{-6}\,\mathrm{kg}$ ), one obtains $D_a \sim 10^{-12}\,\mathrm{m^2/s}$. Using the high-probability return time
\begin{equation}
T_{0.999968} \approx 3\times10^{-4}\,\mathrm s,
\end{equation}
the resulting displacement during one renewal cycle is
\begin{equation}
\Delta a_{\mathrm{RM}} \sim \sqrt{D_a T_{0.999968}}
\sim 10^{-8}\,\mathrm m.
\end{equation}
Over the same interval, the free Schr\"odinger spreading of a Gaussian packet with $\sigma\sim10^{-6}\,\mathrm m$ and $M\sim10^{-6}\,\mathrm{kg}$ is $\Delta\sigma\sim10^{-20}\,\mathrm m$, which is entirely negligible.

Thus, with overwhelming probability, the stochastic deviation accumulated between successive returns to $\widetilde M^{\sigma}_{1,1}$ is on the order of nanometers. Since the diffusion process is reinitialized upon each return, such deviations do not accumulate over macroscopic observation times, and the Newtonian trajectory remains statistically stable even over arbitrarily long intervals, up to exceptionally rare excursions. Exceptional long excursions remain possible, but their probability is comparable to the classical likelihood of finding a particle far from its expected position under a Gaussian distribution, whose tails are non-zero over all space. In this sense, the rare deviations predicted by the {\bf (RM)} dynamics are statistically indistinguishable from ordinary classical fluctuations and do not introduce any new instability beyond that already present in classical probabilistic descriptions.

The obtained behavior is easily visualized on the $(\tau,s)$ plane: the state executes a random walk on $\mathbb{R}^2$ with a Newtonian drift along the $\tau$-axis and a small drift in $s$. The environment acts as a detector array that registers a position whenever the walk enters $s\le 0$ (i.e., when the state lies on $\widetilde M^{\sigma}_{1}$), after which the walk restarts from the recorded value of $\tau$. The sequence of recorded positions $\tau_m$ forms a set of narrow conditional distributions whose widths depend on the time since the previous detection.
Note also that registering the state when it lies on  $\widetilde M^{\sigma}_1$ does not constitute collapse: the state is already reduced, and the detector merely confirms this with certainty.

This reproduces the behavior of a macroscopic body, except that the random walk occurs in a neighborhood of the set underlying $\widetilde M^{\sigma}_{1,1}$ within the full state space, rather than being confined strictly to the classical space $\widetilde M^{\sigma}_1$ or the classical phase space $\widetilde M^{\sigma}_{1,1}$. 
Successive returns to $\widetilde M^{\sigma}_{1}$ then generate a sequence of recorded positions on $\mathbb{R}$ that are normally distributed around the Newtonian trajectory. The Newtonian dynamics of a macroscopic body under these conditions then follows, including the classical behavior of macroscopic measuring devices themselves.

As in Brownian motion on $\mathbb{R}^3$, the time-step and variance parameters of the walk in {\bf (RM)} may vary with the properties of the body, the environment, and their interaction. In deriving Newtonian motion, we argued in favor of many frequent, sufficiently small steps so that the diffusion coefficient $D_a$ is negligible. With appropriate tuning of these parameters, the motion of microscopic particles in various media can be described within the same dynamical framework. In particular, the observed similarity between the tracks of microscopic particles in a bubble chamber and the trajectories of macroscopic bodies in natural environments may reflect comparable parameter regimes.

Conversely, modeling the position measurement of a microscopic particle under typical laboratory conditions requires a very small time step $dt$ but larger step sizes $dz$, and hence a larger diffusion coefficient $D_a$. The resulting random walk spreads into the surrounding state space; however, by symmetry of the walk, after a short time the state still has probability $\approx 1/2$ of lying on $\widetilde M^{\sigma}_{1}$. At such moments, the interaction with the measuring device is effectively Newtonian, producing a classical measurement record. If the measurement interval is sufficiently short, the contribution of the free Hamiltonian $\widehat h$ may be neglected, eliminating the associated drift and yielding the Born rule without Newtonian displacement through the action of $\widehat h_{\mathrm{RM}}$ alone  \cite{KryukovPhysicsA}.
Detailed estimates of parameter values for different systems and environments, as well as an analysis of the influence of these parameters on the motion of microscopic and macroscopic bodies, will be presented in subsequent work. See \cite{KryukovPhysicsA} for an analysis of the double-slit experiment within this framework.

Taken together, the estimates obtained above for the macroscopic case characterize the classical regime in the present model. The inequalities
\begin{equation}
\frac{v\tau}{\sigma}\ll 1,
\qquad
\frac{\tau}{T_{\mathrm{spr}}}\ll 1,
\qquad
P_\sigma[\widehat h,\widehat h_{RM}]P_\sigma = O(\varepsilon),\ \ \varepsilon\ll 1,
\end{equation}
show that during each environmental interaction window the deterministic tangent displacement is observationally negligible, the orthogonal spreading is negligible, and the commutator of the free and {\bf (RM)} contributions is negligible on the dynamically relevant sector. Together with the estimate that the stochastic momentum correction is many orders of magnitude smaller than the classical momentum and with the sub-millisecond return time to the tubular neighborhood of \(M_{1,1}^\sigma\), this implies that the state remains, with overwhelming probability, close to the classical manifold and that its recorded motion is Newtonian on the resolution scale \(\sigma\). Thus the macroscopic classical regime is defined here by a combined drift--diffusion--resolution condition rather than by any single time, mass, or length scale taken in isolation.

We conclude that the same linear Schr\"odinger dynamics, supplemented by the random-matrix term {\bf (RM)}, yields Born-rule state reduction for microscopic particles and Newtonian trajectories for macroscopic ones. The quantum-to-classical transition therefore appears in the present framework as a change of regime within a single dynamical model rather than as a consequence of different underlying laws. In particular, the effective confinement of a macroscopic particle's state to a classical phase-space submanifold is not an added postulate but an emergent feature of the combined drift--diffusion dynamics together with conditioning on detection events.


\section{Comparison with existing approaches}

\subsection{Relation to decoherence theory.}

The modern theory of decoherence (environment-induced superselection, pointer states, collisional decoherence, etc.) provides a well-developed account of suppression of interference and stability of narrow wave packets under environmental monitoring \cite{Mazzola2010,Zurek2003}. For instance, \cite{Fink} explores gradual emergence of classical response as thermal occupation grows, while \cite{Ghose} analyzes the role of entanglement and back-action in continuous monitoring and classical emergence. Experimental studies of quantum-classical crossovers involving light coherence and other observables, e.g.\ \cite{Exp}, show continuous transition regimes in controlled systems. Related studies based on coarse-grained measurements \cite{NaikPanigrahi2024,KoflerBrukner2007,Mukherjee2019,JeongLimKim2014} likewise exhibit controlled quantum-to-classical transition regimes. These works demonstrate how environmental coupling selects robust pointer states and yields classical-like reduced dynamics.

{\bf (RM)} dynamics is fully compatible with this framework at the level of diffusion scales: the parameters entering {\bf (RM)} (e.g., \(D_p\)) are directly traceable to collisional decoherence theory. However, decoherence alone yields an improper mixture in the reduced density matrix; it does not account for the emergence of a single outcome with Born-rule probabilities. The conjecture {\bf (RM)}, combined with equivalence classes, extends this picture by proposing that, under coarse-graining, the effective state evolution becomes isotropic diffusion in projective Hilbert space, yielding genuine single outcomes with Born probabilities while preserving strict unitarity.

A related microscopic perspective is provided by system--environment models such as the Caldeira--Leggett model \cite{CaldeiraLeggett1985}, in which a distinguished degree of freedom is linearly coupled to a bath of harmonic oscillators. Such models derive dissipation and decoherence from an explicit Hamiltonian and lead, after tracing out the bath, to effective reduced dynamics for the system. The role of {\bf (RM)} is different. It is not introduced here as a competing microscopic derivation of environmental coupling, but as an effective stochastic description of the cumulative action of many short and complicated interaction events on the full state space. In this sense, Caldeira--Leggett-type models and {\bf (RM)} operate at different levels: the former provide microscopic bath models, while the latter provides an effective geometric model of the induced diffusion. At the same time, the diffusion scales entering {\bf (RM)} are intended to be compatible with the parameters familiar from decoherence theory, as reflected in the estimates given above.

Another route to classicality, closer in spirit to \cite{Caticha2017,Cui2023}, derives classical center-of-mass motion for large quantum systems through many-particle concentration or central-limit-type arguments. In such approaches, the center-of-mass distribution becomes sharply peaked, often approximately Gaussian, and the resulting trajectory satisfies classical equations of motion. This line of work is relevant to the general problem addressed here, but it differs conceptually from the present framework. In our construction, Gaussian functions are used only as convenient representatives of localized states, viewed as points of the state space where the dynamics takes place; the geometric derivation of classical dynamics does not depend on exact Gaussianity and extends to equivalence classes of arbitrary sufficiently localized states. The essential ingredients here are instead the geometry of the localized phase-space submanifold of state space and, for macroscopic particles, the dynamical stabilization of its neighborhood by the {\bf (RM)} mechanism.

\subsection{Relation to continuous measurement and unravelings.}

In continuous measurement theory (quantum trajectories) \cite{WisemanMilburn,JacobsSteck},
one begins with a Lindblad master equation
\begin{equation}
\dot\rho
=
-\frac{i}{\hbar}[H,\rho]
+
\sum_k \mathcal{D}[L_k]\rho,
\end{equation}
describing the reduced dynamics of a system interacting with an environment.

One then constructs a stochastic unraveling,
leading to a stochastic Schr\"odinger equation (SSE) of the form
\begin{equation}
d|\psi\rangle
=
\left(
-\frac{i}{\hbar}H\,dt
+
\text{drift}
\right)|\psi\rangle
+
\sum_k \text{noise term}\, dW_k .
\end{equation}

The defining structural features are: (i) The conditional state evolution is explicitly stochastic; (ii) The equation is nonlinear in the state; (iii) Noise statistics are chosen so that ensemble averaging reproduces the master equation; (iv) Single outcomes arise due to the nonlinear stochastic terms; (v) The Born rule is built into the construction through consistency with the Lindblad dynamics.

Thus, in quantum trajectory theory,
stochasticity is introduced at the level of the effective state equation.
The stochastic Schr\"odinger equation is postulated as an unraveling
of the master equation.
{\bf (RM)} differs conceptually: the underlying evolution remains unitary, and stochasticity arises from the conjectured random-matrix structure. The Born rule is not imposed through noise statistics but instead follows from the isotropy of the induced diffusion in projective space and the use of equivalence classes.

\subsection{Difference from the Ehrenfest limit.}

The Ehrenfest theorem \cite{Ehrenfest1927} ensures that expectation values approximately follow classical equations of motion when wave packets remain narrow. However: (i) Ehrenfest dynamics applies only to expectation values; (ii) It does not suppress interference; (iii) It does not guarantee stability of wave packets; (iv) It does not produce single measurement outcomes.
Decoherence theory improves upon Ehrenfest by explaining suppression of interference and pointer-state stability. {\bf (RM)} dynamics goes further by directly addressing outcome selection, preserving strict unitarity, explaining Born-rule statistics, and accounting for the transition to Newtonian dynamics.

\subsection{Relation to collapse models}

Collapse models \cite{Bassi2003,Bassi2013} introduce explicit non-unitary stochastic terms into the Schr\"odinger equation, leading to localization events that select outcomes. These models successfully produce single outcomes, but at the price of modifying quantum mechanics and typically introducing universal heating or diffusion.

In contrast, {\bf (RM)} dynamics: (i) preserves strict unitarity; (ii) requires no fundamental modification of the Schr\"odinger equation; (iii) derives stochasticity from geometric diffusion induced by environmental coupling; (iv) produces single outcomes with Born probabilities; (vi) does not imply universal heating beyond environmental effects.

\section{Supplement}

\subsection{Equivalence Classes of localized States and Their Geometry}

Fix $\sigma>0$.  Let $g_c := g_{c,\sigma}$ be the unit Gaussian function in 
$L_2(\mathbb{R})$ of width $\sigma$ centered at $c\in\mathbb{R}$.  
The equivalence class $\{g_c\}\subset L_2(\mathbb{R})$ is defined by
\begin{equation}
\varphi \in \{g_c\}
\quad\Longleftrightarrow\quad
\mu_z(\varphi)=c,\qquad \delta_z(\varphi)\le\sigma.
\end{equation}
Each class $\{g_c\}$ contains infinitely many mutually orthogonal 
elements of $L_2(\mathbb{R})$.

Let $\rho(\psi,\varphi)$ denote the Fubini--Study distance in 
$\mathbb{CP}^{L_2}$.  
The distance from $\psi$ to the class $\{g_c\}$ is defined by
\begin{equation}
\label{distC}
\rho(\psi,\{g_c\}) = \inf_{\varphi\in\{g_c\}} \rho(\psi,\varphi).
\end{equation}

\paragraph*{Criterion for membership.}
A state $\psi$ belongs to $\{g_c\}$ if and only if
\begin{equation}
\rho(\psi,\{g_c\}) = 0.
\end{equation}
Necessity is obvious; sufficiency follows because the class $\{g_c\}$ contains 
a complete family of functions with $\mu_z=c$ and $\delta_z\le\sigma$.

The distance between two classes $\{g_c\}$ and $\{g_d\}$ is defined by
\begin{equation}
\rho(\{g_c\},\{g_d\}) = \inf_{\psi\in\{g_c\}} \rho(\psi,\{g_d\}).
\label{Xeqn80-80}
\end{equation}
To first order of $|c-d|$, this distance coincides with the distance between the Gaussian representatives $g_{c,\sigma}$ and $g_{d,\sigma}$.
Using their inner product, we obtain
\begin{equation}
\cos^2\rho(g_{c,\sigma},g_{d,\sigma})
 = \left|\langle g_{c,\sigma},g_{d,\sigma}\rangle \right|^2
 = \exp\!\left[-\frac{(c-d)^2}{4\sigma^2}\right].
\label{Xeqn81-81}
\end{equation}
Thus
\begin{equation}
\rho(g_{c,\sigma},g_{d,\sigma})
  = \arccos\!\left( e^{-(c-d)^2/8\sigma^2} \right).
\label{Xeqn82-82}
\end{equation}

Under the embedding into $\mathbb{CP}^{L_2}$, the set
\begin{equation}
M_1^\sigma = \{ g_{a,\sigma}\  \vert\ a\in\mathbb{R} \}
\label{Xeqn83-83}
\end{equation}
is a one-dimensional submanifold with the induced Fubini--Study metric.
With an appropriate choice of units, the map $a\mapsto g_{a,\sigma}$ is an
isometry between the Euclidean line $\mathbb{R}$ and $M_1^\sigma$.
Using Gaussian representatives $g_{c,\sigma}$ to define distances, the set
\begin{equation}
\widetilde M_1^\sigma = \{ \{g_a\}\ \vert\ a\in\mathbb{R} \}
\label{Xeqn84-84}
\end{equation}
with the resulting metric is likewise a Riemannian manifold isometric to both $M_1^\sigma$ and $\mathbb{R}$.

Similarly,
\begin{equation}
M^{\sigma}_{1,1}
  = \{ g_{a,\sigma}(z)\, e^{ipz/\hbar}\ \vert\ a,p\in\mathbb{R} \}
\label{Xeqn85-85}
\end{equation}
embeds as a two-dimensional submanifold of $\mathbb{CP}^{L_2}$, and the
corresponding set of equivalence classes
\begin{equation}
\widetilde M^{\sigma}_{1,1}
  = \big\{\, \{g_{a}\}\cdot e^{ipz/\hbar}\ \vert\  a,p\in\mathbb{R} \,\big\}
\label{Xeqn86-86}
\end{equation}
is defined analogously.
Distances defined using Gaussian representatives satisfy
\begin{equation}
\cos^2 \rho\!\left(g_{a,\sigma} e^{ipz/\hbar},\,
                  g_{b,\sigma} e^{iqz/\hbar}\right)
 =
 \exp\!\left[-\tfrac{(a-b)^2}{4\sigma^2}
             -\tfrac{(p-q)^2\sigma^2}{\hbar^2}\right].
\label{Xeqn87-87}
\end{equation}
With suitable units, this yields the isometries
\begin{equation}
\widetilde M^{\sigma}_{1,1}\;\cong\; M^{\sigma}_{1,1}\;\cong\; \mathbb{R}^2.
\label{Xeqn88-88}
\end{equation}

\subsection{Finite Gaussian Partition Spaces and Orthogonality of $(\tau,s)$ Coordinates}

Let $z_1,\dots,z_N$ be an equally spaced finite set of points in $\mathbb{R}$, and
define equal-width Gaussian functions centered at these points
\begin{equation}
g_k(z) := g_{z_k,\sigma}(z).
\end{equation}
Assume that the width $\sigma$ is much smaller than the spacing, so
that the functions are approximately orthogonal.  
Define the finite-dimensional subspace of $L_2(\R)$
\begin{equation}
V = \mathrm{span}\{ g_1,\dots,g_N\}.
\end{equation}

The set of all finite linear combinations of translates of a single Gaussian function is
dense in $L_2(\mathbb{R})$ \cite{KryukovMath,Gaussians}.  
Moreover, any function in $L_2(\mathbb{R})$ can be approximated to arbitrary precision by a finite superposition of approximately orthogonal, equal-width Gaussians of sufficiently small width. In fact, this is true for continuous functions, which form a dense subset of $L_2(\mathbb{R})$.
Because of finite detector resolution, measurement outcomes are naturally described in terms of equivalence classes of states that correspond to equivalent probability distributions of the particle's position as measured by the detector. Thus, any experimentally distinguishable state can be represented by a vector in a corresponding space $V$.

For any $\varphi\in V$, define
\begin{equation}
\varphi_{\tau,\lambda}(z)
  = \sqrt{\lambda}\,\varphi\!\big(\lambda(z-\mu_z-\tau)+\mu_z\big),
\end{equation}
and set $s=\ln(\lambda/\sigma)$.  
A displacement in $\tau$ shifts the expectation value of position, whereas a
displacement in $s$ rescales the spread $\delta_z$ \cite{KryukovPhysicsA}.
Since $\mu_z$ and $\delta_z$ are independent of the global phase of $\varphi$,
this construction defines a two-dimensional submanifold $M_\varphi$ of the
projective space $\mathbb{CP}(V)$.

Assume $\varphi \in V$ and $||\varphi||=1$. 
Then,
\begin{equation}
\int \varphi(z)\,\varphi'(z)\,dz = 0
\end{equation}
holds exactly, and
\begin{equation}
\int (z-\mu_z)\,|\varphi'(z)|^2\,dz = 0,
\end{equation}
up to exponentially small corrections.
These relations imply orthogonality of the tangent vectors $\partial_\tau \varphi_{\tau,s}$ and $\partial_s \varphi_{\tau,s}$:
\begin{equation}
\left\langle \partial_\tau \varphi_{\tau,s},\,\partial_s \varphi_{\tau,s}
\right\rangle = 0.
\end{equation}
It follows that $(\tau,s)$ form an orthogonal coordinate system on the two-dimensional
submanifold
\begin{equation}
M_\varphi := \{\, \varphi_{\tau,s} \left. \right | (\tau,s)\in\mathbb{R}^2 \,\} \subset  \mathbb{CP}^{L_2},
\end{equation}
 which is therefore isometric to the Euclidean plane $\mathbb{R}^2$.  
Moreover, the manifolds $M_\varphi$ arise as orbits of the action of the
translation group in $\tau$ and $s$ on the dense subspace of $L_2(\mathbb{R})$
consisting of functions with finite $\mu_z$ and $\delta_z$.

\subsection{State reduction as a motion on $\R^2$}

Reduction or collapse of the state under a position measurement, as described 
by Schr{\"o}dinger evolution with the Hamiltonian in {\bf (RM)}, is understood 
as the approach of the state to the manifold $\widetilde M^{\sigma}_1$ in the 
metric~(\ref{distC}).

For each point of $M_\varphi$, specified by coordinates $(\tau,\lambda)$ and 
represented by an element of the unit sphere $S(V)\subset V$, consider the set 
of all functions in $S(V)$ satisfying $\mu_z=\tau$ and $\delta_z=\lambda$.  
These sets form a foliation of $S(V)$ of codimension two; each leaf 
$\{\varphi\}_{\tau,\lambda}$ consists of all states in $S(V)$ sharing the same 
values of $\mu_z$ and $\delta_z$.

More formally, define the smooth map
\begin{equation}
F : S(V) \longrightarrow \mathbb{R}^2,
\qquad 
F(\varphi) = \big(\mu_z(\varphi),\,\delta_z(\varphi)\big).
\end{equation}
The map $F$ has full rank on $S(V)$ except on a subset $S_{\mathrm{sing}}(V)$ 
of measure zero.  
On the regular set $S_{\mathrm{reg}}(V):=S(V)\setminus S_{\mathrm{sing}}(V)$, 
the level sets $F^{-1}(\mu_z,\delta_z)$ are smooth submanifolds of codimension~$2$, 
and these leaves fit together to form a smooth foliation of $S_{\mathrm{reg}}(V)$.  
Thus, each generic state lies on a unique leaf specified by its mean position and spread, while nongeneric states, for which the rank of $dF_\varphi$ drops, form a thin exceptional set that does not affect the reduction dynamics, as they can be well approximated by generic states.  
Since $F(e^{i\theta}\varphi)=F(\varphi)$, the same construction produces a foliation 
of the projective space $\mathbb{CP}(V)=S(V)/S^1$.

Because $(\tau,s)$ are orthogonal coordinates on $M_\varphi$ for each 
$\varphi\in\mathbb{CP}(V)$, the corresponding components of a step of the random 
walk in {\bf (RM)} are independent random variables.  
Since the step distribution in {\bf (RM)} is homogeneous, the probability laws 
for the $\tau$- and $s$-components, measured in the Fubini-Study metric, are 
identical at all points $\varphi$.  
By definition, $\mu_z$ and $\delta_z$ remain constant along the leaves through 
$M_\varphi$.  
Consequently, step components tangent to a leaf do not change $\mu_z$ or 
$\delta_z$, and therefore do not contribute to collapse into a physical eigenstate 
of position.

Thus, in this framework, state reduction can be fully described within the 
two-dimensional space $M_\varphi \cong \mathbb{R}^2$.  
In particular, after sufficiently many steps, the initial state is guaranteed to reach an equivalence class
$\{g_c\}$.

\section*{Acknowledgements}
The author thanks the reviewers for drawing attention to additional literature on decoherence, coarse-grained measurements, and center-of-mass classicalization that helped improve the discussion of related approaches.

\section{Declaration of interest statement}

The author declares no competing interests.


\begin{thebibliography}{99}

\bibitem{Donadi2021}S.~Donadi \emph{et al.}, 
Eur. Phys. J. C \textbf{81}, 773 (2021).

\bibitem{Piscicchia2024}K.~Piscicchia \emph{et al.},
Phys. Rev. Lett. \textbf{132}, 250203 (2024).

\bibitem{Bilardello2016}M.~Bilardello \emph{et al.},
Phys. Rev. A \textbf{94}, 042124 (2016).

\bibitem{Vinante2015}A.~Vinante \emph{et al.},
Phys. Rev. Lett. \textbf{116}, 090402 (2016).

\bibitem{Carlesso2016}M.~Carlesso \emph{et al.},
Phys. Rev. D \textbf{94}, 124036 (2016).

\bibitem{Altamura2025}E.~Altamura \emph{et al.},
Phys. Rev. A \textbf{111}, L020203 (2025).

\bibitem{Adler2007}S.~L.~Adler,
J. Phys. A \textbf{40}, 2935 (2007).

\bibitem{AdlerBassi2009}S.~L.~Adler and A.~Bassi,
Science \textbf{325}, 275--276 (2009).

\bibitem{IGMHeating2017}V.~Mukhanov \emph{et al.},
Phys. Rev. D \textbf{95}, 063505 (2017).

\bibitem{KryukovPhysicsA}A.~Kryukov,
J. Phys. A: Math. Theor. \textbf{58}, 225302 (2025).

\bibitem{Kryukov2020}A.~Kryukov,
J. Math. Phys. \textbf{61}, 082101 (2020).

\bibitem{KryukovNew}A.~Kryukov,
J. Phys.: Conf. Ser. \textbf{2482}, 012025 (2023).

\bibitem{KryukovDICE24}A.~Kryukov,
J. Phys.: Conf. Ser. \textbf{3017}, 012018 (2024).

\bibitem{KryukovMath}A.~Kryukov,
Int. J. Math. Math. Sci. \textbf{2005}, 2241 (2005).

\bibitem{delta}J.~M.~Aguirregabiria, A.~Hernandez, and M.~Rivas,
Am. J. Phys. \textbf{70}, 180--185 (2002).

\bibitem{KryukovPhysLett}A.~Kryukov,
Phys. Lett. A \textbf{370}, 419 (2007).

\bibitem{Ein}A.~Einstein,
Ann. Phys. (Leipzig) \textbf{322}, 549--560 (1905).

\bibitem{Wigner}E.~P.~Wigner,
Proc. Cambridge Philos. Soc. \textbf{47}, 790 (1951).

\bibitem{BGS}O.~Bohigas, M.~J.~Giannoni, and C.~Schmit,
Phys. Rev. Lett. \textbf{52}, 1 (1984).

\bibitem{Mazzola2010}
L.~Mazzola, J.~Piilo, and S.~Maniscalco,
Phys. Rev. Lett. \textbf{104}, 200401 (2010).

\bibitem{Zurek2003}
W.~H.~Zurek,
Rev. Mod. Phys. \textbf{75}, 715 (2003).


\bibitem{Fink}J.~M.~Fink \emph{et al.},
Phys. Rev. Lett. \textbf{105}, 163601 (2010).

\bibitem{Ghose}S.~Ghose \emph{et al.},
Phys. Rev. A \textbf{72}, 014102 (2005).

\bibitem{Exp}P.~Lassegues \emph{et al.},
Phys. Rev. A \textbf{108}, 042214 (2023).


\bibitem{NaikPanigrahi2024}
L.~P.~Naik and K.~Panigrahi,
Phys. Rev. A \textbf{109}, 022202 (2024).

\bibitem{KoflerBrukner2007}
J.~Kofler and \v{C}.~Brukner,
Phys. Rev. Lett. \textbf{99}, 180403 (2007).

\bibitem{JeongLimKim2014}
H.~Jeong, Y.~Lim, and M.~S.~Kim,
Phys. Rev. Lett. \textbf{112}, 010402 (2014).

\bibitem{Mukherjee2019}
S.~Mukherjee, A.~Rudra, D.~Das, S.~Mal, and D.~Home,
Phys. Rev. A \textbf{100}, 042114 (2019).

\bibitem{CaldeiraLeggett1985}
A.~O.~Caldeira and A.~J.~Leggett,
Phys. Rev. A \textbf{31}, 1059 (1985).

\bibitem{Cui2023}
B.~Cui,
Phys. Lett. A \textbf{482}, 129041 (2023).

\bibitem{Caticha2017}
A.~Demme and A.~Caticha,
AIP Conf. Proc. \textbf{1853}, 090001 (2017).


\bibitem{WisemanMilburn}
H.~M.~Wiseman and G.~J.~Milburn,
\emph{Quantum Measurement and Control}
(Cambridge University Press, Cambridge, 2010).

\bibitem{JacobsSteck}
K.~Jacobs and D.~A.~Steck,
Contemp. Phys. \textbf{47}, 279 (2006).

\bibitem{Ehrenfest1927}
P.~Ehrenfest,
Z. Phys. \textbf{45}, 455 (1927).

\bibitem{Bassi2003}
A.~Bassi and G.~C.~Ghirardi,
Phys. Rep. \textbf{379}, 257 (2003).

\bibitem{Bassi2013}
A.~Bassi, K.~Lochan, S.~Satin, T.~P.~Singh, and H.~Ulbricht,
Rev. Mod. Phys. \textbf{85}, 471 (2013).

\bibitem{Gaussians}
C.~Calcaterra and A.~Boldt,
arXiv:0805.3795 (2008).

\end{thebibliography}
\end{document}